\documentstyle[preprint,aps]{revtex}
\begin{document}

\title{Independent domains of disoriented chiral dondensate} 

\author{Q. H. Zhang} 
\address{Instit\"ut f\"ur Thoretisch Physik,
Universit\"at Regensburg,
D-93040 Regensburg, Germany}
\author{X. Q. Li}
\address{China Center of Advanced Science and Technology(World Laboratory),
 P.O. Box 8730, Beijing 100080, P.R. China\\
Physics Department, Nankai University, Tianjing 300071, China
}

\maketitle

\begin{abstract}
The probability distribution of a neutral pion fraction from 
independent domains of disoriented chiral condensate is 
characterized. 
The signal for the condensate is still clear for a large number 
of independent domains if one of them is predominant.


\end{abstract}

A disoriented dhiral condensate (DCC) may be formed in large hot regions of 
hadronic matter where the approximate chiral symmetry of QCD has been briefly restored. 
The decay of this region induces a non-equilibrium relaxation of the chiral 
fields which is predicted to create coherent sources of soft pion modes
\cite{Raj92}.  The charge of the pions emitted from the DCC has a characteristic 
probability distribution\cite{BK92,Bjo93}
\begin{equation}
P(f)=\frac{1}{2\sqrt{f}}
\end{equation}
where 
\begin{equation}
f=\frac{n_0}{n_{total}}.
\end{equation}
Here $n_0$ is the number of observed neutral pions while $n_{total}$ is the 
total number of observed pions. 
This is markedly different from the standard  statistical distribution 
which, for large $n_{total}$, we expect to be
\begin{equation}
P(f) = \delta (f-\frac{1}{3}).
\end{equation}
Using the discrete wavelet transformation method, 
Huang et al.\cite{HSTW96}
calculated the probability distribution of the neutral pion fraction $f$ 
for different physical scales and found that due to 
the DCC cluster size, $P(f)$ exhibits a delay in approaching Eq.(3) 
required by the central limit theorem ({\it delayed central limit}). 
This can be used as a signature of DCC in high energy heavy-ion collisions. 

Recently Amado and Lu\cite{AL96} claimed that for more than 
three independent domains of DCC the signal for condensate (Eq.(1)) 
will be reduced. But their conclusion is based on the assumption that 
the $N$ independent regions have 
equal weight which means that each region emits the same total number of 
pions. In the case that 
the number of the regions goes 
to infinity, the probability distribution of the neutral pion fraction becomes 
$P(f)=\delta(f-1/3)$. In this Brief Report, we will prove that 
if one of these regions is predominant,  that is, if most pions are emitted 
from one of the uncorrelated domains, we can still have a clear 
signal for the condensate 
even though the number of the independent regions is very large. 
It is likely that, particularly in a heavy ion
collisions, more than one domain or
 region of chiral condensate
will be formed in the large interaction volume\cite{HSTW96}.  Similar to ref.\cite{AL96},
we assume that these domains are uncorrelated, that is,
pions are emitted independently from each domain.  In that 
case the observed neutral pion fraction $f$ will be an average
 over these independent regions. It is also likely that 
each region has a different weight 
$\alpha_{i}=\frac{n_{i,total}}{\sum_{i} n_{i,total}}$.
Here $n_{i}$ is the total number of pions emitted from region $i$. 
In each single region we take the probability 
of $f$ to be given by $P_1(f)$ (eq.(1)).  Then the probability of 
finding a neutral fraction $f$ averaged over the N 
regions is given by
\begin{equation}
P_N(f,\alpha_i) = \int df_1 ....df_N \delta\left(f-\sum_{i=1}^{N}\alpha_{i}f_{i}
\right)
    P_1(f_1) P_1(f_2).....P_1(f_N).
\end{equation}
This can be transformed into a recursion relation,
\begin{equation}
P_N(f,\alpha_i) = \frac{1}{1-\alpha_{1}} \int P_{N-1} 
\left(\frac{f -\alpha_{1}f_1}{1-\alpha_{1}},\frac{\alpha_{i}}{1-\alpha_{1}}\right) 
P_1(f_1) d f_1.
\end{equation}
This relation is particularly helpful in computing $P_N$ stepwise in N.
For two domains the probability
can be found analytically and we get
\begin{equation}
P_2(f,\alpha_{i}) = \frac{\pi}{4\sqrt{\alpha_{1}\alpha_{2}}}
\end{equation} 
for $f<min\{\alpha_{1},\alpha_{2}\}$, 
\begin{equation}
P_2(f,\alpha_{i}) = \frac{1}{4\sqrt{\alpha_{1}\alpha_{2}}}
\left[\pi -2 \arccos (\sqrt{\frac{min\{\alpha_{1},\alpha_2\}}{f}})\right]
\end{equation} 
for $min\{\alpha_1,\alpha_2\}\le f \le max\{\alpha_1,\alpha_2\}$ and
\begin{equation}
P_2(f,\alpha_{i}) = \frac{1}{4\sqrt{\alpha_{1}\alpha_{2}}}
\left[\pi -2 \arccos (\sqrt{\frac{\alpha_{1}}{f}})-
 2 \arccos (\sqrt{\frac{\alpha_{2}}{f}})\right]
\end{equation} 
for $f>max\{\alpha_1,\alpha_2 \}$. For the case $\alpha_1=\alpha_2=1/2$ the 
result was previously calculated in \cite{HSTW96,AK95}.  
In the following, we will show that for $\alpha_{1}=
\alpha_{2}=\cdot \cdot \cdot=\alpha_{N}=1/N $, as
$N $ tends to infinity, the probability distribution reaches exactly eq.(3).  
It is easily checked that eq.(4) can be written down as

\begin{eqnarray}
G(f)&=&\lim_{N \rightarrow \infty}P_{N}(f)
\nonumber\\
&=&\frac{1}{2\pi}\lim_{N \rightarrow \infty}\int df_{1}df_2\cdot \cdot \cdot
df_N dy e^{ify-i\frac{f_1+\cdot \cdot \cdot + f_N}{N}y} P_1(f_1)\cdot \cdot 
\cdot P_1(f_N)\\
&=&\frac{1}{2\pi}\lim_{N \rightarrow \infty}\int df_{1}df_2\cdot \cdot \cdot
df_N dy e^{ify}\sum_{n=0}^{\infty}(-i)^n \frac{(f_1+\cdot\cdot\cdot +f_N)^n}
{N^n n!} P_1(f_1)\cdot \cdot 
\cdot P_1(f_N)  
~~,   
\nonumber
\end{eqnarray}
Using the relationship that 
\begin{equation}
\lim_{N \rightarrow \infty}\int df_{1}df_2\cdot \cdot \cdot
df_N  \frac{(f_1+\cdot \cdot \cdot + f_N)^n}{N^{n}n!}
P_1(f_1)\cdot \cdot \cdot P_1(f_N)=\frac{1}{3^n n!}  ,
\end{equation}
we have 
\begin{equation}
G(f)=\delta(f-1/3)   .
\end{equation}
Now we consider the non equal-weight case and for simplicity we assume that 
$\alpha_1=0.8$, that is, eighty percent of the pions are emitted from one of those 
independent domains, the
other pions are emitted by other regions with weight
 $\alpha_{2}=\cdot \cdot \cdot =\alpha_N =0.2/(N-1)$. The numerical results are shown in 
Fig.1 . For $N \rightarrow \infty$, we have
\begin{equation}
G(f)=0
\end{equation}
for $f \le 1/15$
\begin{equation}
G(f)=\frac{5}{4}\frac{1}{\sqrt{5f-1/3}}
\end{equation}
for $1/15 \le f \le 13/15$
and 
\begin{equation}
G(f)=0
\end{equation}
for $13/15 \le f$. This shows clearly the difference between our results and 
previous results \cite{AL96}. Even with many independent regions of DCC, 
we can still have a clear signal of DCC if one of these regions is 
predominant. 

Conclusion: In this Brief report, we calculated the probability 
distribution of $P(f)$ for more than three independent domains with different 
weights. It was shown that if one of this DCC regions is 
predominant, we can still see the signal of the condensate 
even if the number of domains is very large. 

 One of the authors (Q.H.Z.) 
thank Dr. Stefan Ochs for reading the manuscript.  
This work was supported in part by the Alexander von Humboldt foundation 
in Germany and the National Natural Science Foundation of China. 

\section*{Figure Captions}
   
Fig.1   The probability density $P(f)$ vs.  neutral pion fraction $f$.
\end{document}